\documentclass[amsmath,amssymb,twocolumn,amsfonts]{revtex4}
\usepackage{graphicx}
\def \beq {\begin{equation}}
\def \eeq {\end{equation}}
\def \tr {\rm Tr}

\begin{document}
\title{Second Comment on "Spin-selective reactions of radical pairs act as quantum measurements" (Chemical Physics Letters 488 (2010) 90-93)
}
\author{Iannis K. Kominis}

\affiliation{Department of Physics, University of Crete, Heraklion
71103, Greece}

\maketitle
Spin-selective radical-ion-pair reactions are at the core of spin chemistry \cite{steiner}. Jones \& Hore
recently introduced a master equation, which the authors claim to follow from quantum measurement theory considerations \cite{JH} and which is supposed to describe the evolution of the spin density matrix of radical-ion pairs. 

We have recently shown \cite{kom_comment1} that the Jones-Hore theory is inconsistent, since it cannot unambiguously 
account for the state change of unrecombined radical-ion pairs. In order to remedy the inconsistency, the 
authors of the original paper together with Maeda and Steiner \cite{reply} have by hand expanded the density matrix 
$\rho$ describing radical-ion pairs to a new density matrix $\rho'$ including the neutral reaction products. By doing so, the authors state
that in \cite{kom_comment1}  I incorrectly assumed that $w_P$, the weight of the products, is zero. 

Now, however, the authors face a more daunting 
challenge, namely the description of a single radical-ion pair. The physical question to be addressed is this: Consider a {\bf single} radical-ion
pair at time $t=0$ in the state $\rho_{0}$. Assume that in a given realization of the experiment the radical-ion pair has not recombined until time $t$. What is $\rho_{t}$ ?

In this case there are no product molecules, so the authors in \cite{reply} cannot use the same mathematical trick (essentially the authors in \cite{reply} have proved the tautology $\rho/\tr\{\rho\}=\rho/\tr\{\rho\}$). So now, based on the reply \cite{reply}, the possible conclusions that follow are only these two:
(i) either the Jones-Hore theory is inconsistent, for the reasons outlined in \cite{kom_comment1}, or (ii) the 
Jones-Hore theory cannot describe the state evolution of single molecules. Either of the two renders the theory highly problematic. 
 
To elaborate, the authors in \cite{reply} might insist that the master equation describing unrecombined radical pairs (either a single one or more) is equation (2) of \cite{reply}, which is reproduced here 
\beq
d\rho/dt=-k_{S}(\tr\{Q_{T}\rho Q_{T}\}\rho-Q_{T}\rho Q_{T})\label{eq1}
\eeq
Now, however, there are two problems. If one considers an experiment with a single radical pair, the authors will not be able to 
derive \eqref{eq1} from the average of elemental trajectories, as now recombination is not an option. Hence the authors will run into the
inconsistency analyzed in \cite{kom_comment1}. If the authors postulate \eqref{eq1} without deriving it, they will be describing a {\it single unrecombining} radical pair with a non-linear equation. Either of the two renders the theory highly problematic.

\end{document}